# Close expressions for Meyer Wavelet and Scale Function

Victor Vermehren Valenzuela[1] and H. M. de Oliveira[2]
[1]State University of Amazon, Av. Darcy Vargas 1200, Manaus, AM, Brazil
[2]Federal University of Pernambuco, Statistics Department, Recife, PE, Brazil.

**ABSTRACT**

Many continuous wavelets are defined in the frequency domain and do not have analytical expressions in the time domain. Meyer wavelet is ordinarily defined in this way. In this note, we derive new straightforward analytical expressions for both the wavelet and scale function for the Meyer basis. The validity of these expressions is corroborated by numerical computations, yielding no approximation error.

**Keyterms** – Meyer wavelets, multiresolution analysis, close expressions for Meyer wavelet.

## I. INTRODUCTION

Wavelets were introduced in the eighties by J. Morlet in the context of analysis of seismic signals [1]. The wavelet theory quickly became the outcome of a multidisciplinary endeavor. Particularly, they have lately gained prolific applications throughout an amazing number of areas, especially in Engineering [2,3]. Essentially, the Wavelet Transform is signal decomposition onto a set of basis function, which is derived from a single prototype wavelet by scaling (dilatations and contractions) as well as translations (shifts). The orthogonality has long been assumed as a fundamental property in virtually all standard approaches when analyzing or synthesizing signals. Both Continuous and Discrete Wavelet transforms (CWT and DWT, respectively) have emerged as a definitive tool of signal processing analysis and have proven to be more powerful than classical Fourier analysis in countless situations [2,4]. Continuous wavelet transform furnishes precise information on the local and global irregularities [5] and is very robust against spurious contamination of the signal. The Meyer wavelet has been used in various scenarios, such as in multi fault classification [6], or in adaptive filter [7] and fractal random fields [8]. However, wavelet algorithm for image deblurring is certainly the most attractive advantage offered by wavelet in modern multimedia schemes [9], among many other applications. Other application that requires the knowledge of the analytical expression of the wavelet is wavelet-based OFDM systems [10-12]. Meyer wavelets were also successfully applied as a basis for atomic and molecular systems for solving radial Schrödinger equations for atoms [13]. A number of continuous wavelets do have analytical close expression in time and frequency domain, such as Shannon (sinc), Mexican hat, beta etc. [3, 4]. Recently approximate expressions to represent compactly supported wavelets such as daublets and symlets were introduced [14]. However, an astonishing, but well established feature of wavelets is that in many cases the signal analysis can be carried out without knowing the waveform, i.e. without the time-domain analytical expression of the mother wavelet. This fact is particularly true for the most well-known compactly supported wavelets (db, coif, sym…) and also the Meyer wavelet.

## II. CLOSE EXPRESSIONS FOR MEYER WAVELET AND SCALE FUNCTION

A landmark in the development of wavelets is in 1985 [15], when Y. Meyer, a harmonic analyst, pointed out the strong connection with the existing analysis techniques of singular integral operators and proposed the first nontrivial orthogonal wavelet basis. Unlike the Haar wavelets, they are continuously differentiable, yet they do not have compact support. Meyer together with Mallat also introduced the concept of multiresolution in 1988 [2]. Meyer was awarded with the 2010 Gauss Prize [16]. The Meyer wavelet is an orthogonal wavelet that is indefinitely differentiable with infinite support. The Mayer scale function and wavelet are defined in the frequency domain in terms of function $\upsilon$ by means of well-known equations [17]:

$$\Phi_{mey}(w) = \begin{cases} \frac{1}{\sqrt{2\pi}} & \text{if } w \leq \frac{2\pi}{3} \\ \frac{1}{\sqrt{2\pi}} \cos\left(\frac{\pi}{2} \upsilon(\frac{3w}{2\pi} - 1)\right) & \text{if } \frac{2\pi}{3} \leq w \leq \frac{4\pi}{3} \\ 0 & \text{otherwise} \end{cases} \qquad (1)$$

where, for instance (other choice can be made),

$$\upsilon(x) = \begin{cases} 0 & \text{if } x < 0 \\ x & \text{if } 0 \leq x \leq 1 \\ 1 & \text{if } x > 1 \end{cases}$$

and the wavelet spectrum is given by

$$\Psi_{mey}(w) = \begin{cases} \frac{1}{\sqrt{2\pi}} \sin\left(\frac{\pi}{2} \upsilon(\frac{3|w|}{2\pi} - 1)\right) e^{\frac{jw}{2}} & \text{if } \frac{2\pi}{3} \leq |w| \leq \frac{4\pi}{3} \\ \frac{1}{\sqrt{2\pi}} \cos\left(\frac{\pi}{2} \upsilon(\frac{3|w|}{4\pi} - 1)\right) e^{\frac{jw}{2}} & \text{if } \frac{4\pi}{3} \leq |w| \leq \frac{8\pi}{3} \\ 0 & \text{otherwise} \end{cases} \qquad (2)$$

Their corresponding wave plots are shown in Figures 1 and 2, respectively.

In order to evaluate the corresponding wave forms of the Equations (1) and (2) in time domain, denoted by $\phi_{mey}(t)$ and $\psi_{mey}(t)$, we use the inverse Fourier transform resulting in:

$$\phi_{mey}(t) = \frac{2}{\sqrt{2\pi}} \int_0^{\frac{4\pi}{3}} \Phi_{mey}(w) \cos(wt) dw. \qquad (3)$$

$$\psi_{mey}(t) = 2 \int_{\frac{2\pi}{3}}^{\frac{8\pi}{3}} \Phi_{mey}(\frac{w}{2}) \Phi_{mey}(w - 2\pi) \cos(w(t - 0.5)) dw. \qquad (4)$$

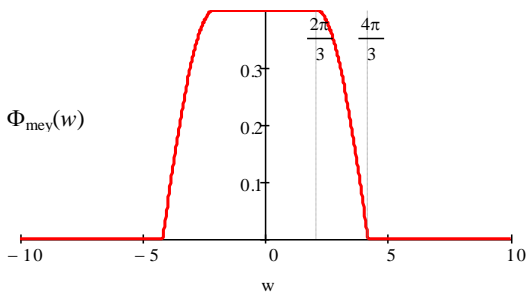

Figure 1. Spectrum of the Meyer scale function. Its support is within spectral band [0, 4π/3].

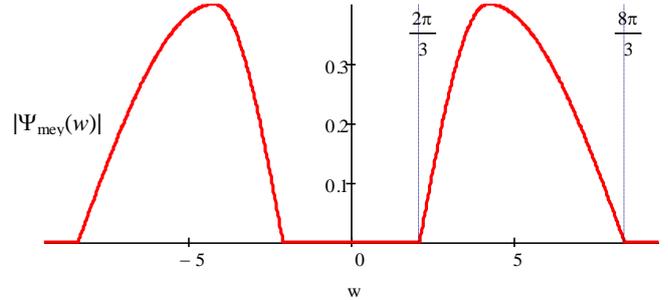

Figure 2. Meyer wavelet curve in the frequency domain. Central frequency is 5π/3. The support is within spectral band [2π/3, 8π/3]

A synchronous detection of the bandpass signal (Meyer wavelet) with a carrier in $w_0 = 6\pi/3$ can be achieved. The detection is done in two parts, "in phase" and "in quadrature" using low-pass filters for the respective components. Figure 3 illustrates the process of wavelet decomposition. The analysis is conveniently separated into upper and lower branches.

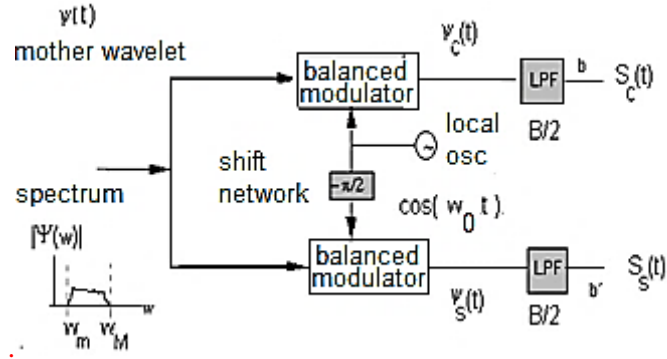

Figure 3. Meyer wavelet decomposition in their low-frequency components in phase and in quadrature.

Figure 3 shows two analysis branches:

a) *Upper Branch*
$$\psi_c(t) := \psi(t).\cos[w_0 t]. \qquad (5a)$$

b) *Lower Branch*
$$\psi_s(t) := \psi(t).\sin[w_0 t]. \qquad (5b)$$

For implementation in filter bank, see [18]. Thus, the band-pass representation yields:

$$S_C(w) = \frac{1}{2}\left(\Psi_{MEY}(w+\frac{6\pi}{3}) + \Psi_{MEY}(w-\frac{6\pi}{3})\right) \quad \text{and} \quad S_S(w) = \frac{j}{2}\left(\Psi_{MEY}(w+\frac{6\pi}{3}) - \Psi_{MEY}(w-\frac{6\pi}{3})\right). \qquad (6)$$

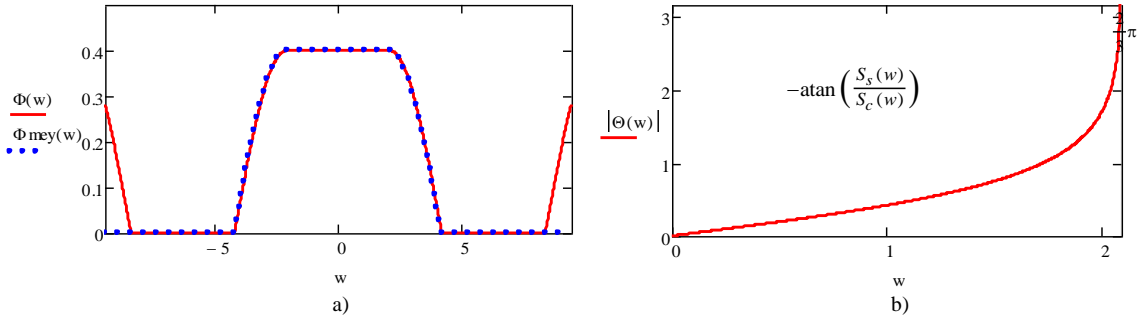

Figure 4. a) Spectrum components in the baseband spectrum for the Meyer wavelet. The low-pass must be selected featuring a limited bandwidth signal at 0.5 Hz  b) Representation of the spectrum phase in terms of the components $S_c$ and $S_s$, cf. Eqn (6).

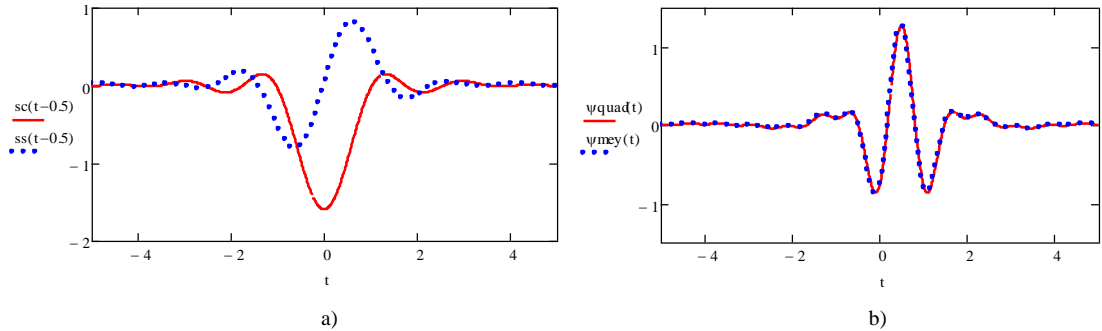

Figure 5. a) Phase and quadrature components of the representation of Meyer wavelet. b) Generating wavelet Meyer from the phase and quadrature components according to the representation of band-pass of Eqn (7). The standard function is obtained from Eqn (4).

From Fig. 3 it is possible to retrieve $\psi_{mey}(t)$ from its components $s_c(t)$ and $s_s(t)$:

$$\psi_{quad}(t) = s_c(t) \cdot \cos(2\pi t) + s_s(t) \cdot \sin(2\pi t) \ . \tag{7}$$

Curiously, the "Lower Side Band" version of $\psi_{mey}(t)$ modulated with a carrier $6\pi/3$, retrieved the same spectrum as the scaling function, $\phi_{mey}(t)$ i.e.

$$\phi_{mey}(t) = \psi_{mey}(t) \cdot \cos(2\pi t) + \hat{\psi}_{mey}(t) \cdot \sin(2\pi t), \tag{8}$$

where $\hat{\psi}_{mey}(t)$ is the Hilbert transform [2] of $\psi_{mey}(t)$.

The envelope of the Meyer scale funcion $\phi_{mey}(t)$ is given by $\sqrt{\psi_{mey}^2(t) + \hat{\psi}_{mey}^2(t)}$. This is the same envelope of the analytical signal $\psi_{mey}(t) + j\hat{\psi}_{mey}(t)$.

Solving the previous integrals with the aid of "tables of integrals" [19-21], it was also possible to derive the following analytical expressions:

- closed analytical expression for the scaling function of Meyer, $\phi_{mey}(t)$,

$$\phi_{mey}(t) = \begin{cases} \left(\frac{2}{3} + \frac{4}{3\pi}\right) & if \ t=0 \\ \dfrac{\sin\left(\frac{2\pi}{3}t\right) + \frac{4}{3}t\cos\left(\frac{4\pi}{3}t\right)}{\pi t - \frac{16\pi}{9}t^3} & otherwise \end{cases} . \tag{9}$$

- closed analytical expression for the Meyer wavelet, $\psi_{mey}(t)$,

$$\psi_1(t) = \frac{\frac{4}{3\pi}(t-0.5)\cos\left[\frac{2\pi}{3}(t-0.5)\right] - \frac{1}{\pi}\sin\left[\frac{4\pi}{3}(t-0.5)\right]}{(t-0.5) - \frac{16}{9}(t-0.5)^3} \tag{10}$$

$$\psi_2(t) = \frac{\frac{8}{3\pi}(t-0.5)\cos\left[\frac{8\pi}{3}(t-0.5)\right] + \frac{1}{\pi}\sin\left[\frac{4\pi}{3}(t-0.5)\right]}{(t-0.5) - \frac{64}{9}(t-0.5)^3} \tag{11}$$

$$\psi_{mey}(t) = \psi_1(t) + \psi_2(t) \tag{12}$$

Analitical similar expressions were not found in the literature. Note that the envelope of the Meyer wavelet decays with $O(t^{-3})$, which is faster than the (*long*) sinc wavelet that decays with $O(t^{-1})$. To check then the validity of the obtained formulae, the integrals corresponding to Equations (3) and (4) were numerically evaluated with the aid of MathCad$^{TM}$ and compared with the results from Equations (9) and (12), respectively. No numerical error was found. This can also be observed in Figures 6 and 7 in the sequel.

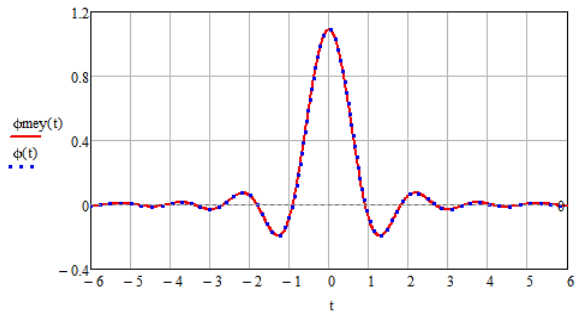
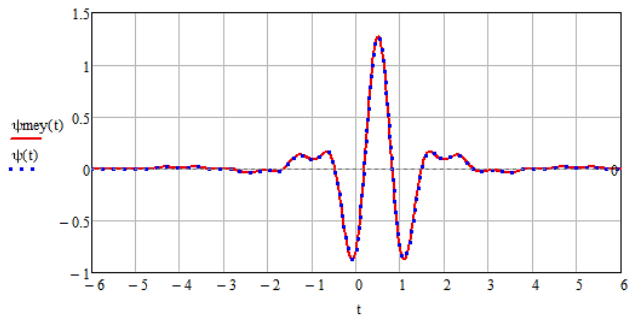

Figure 6. Scale function of Meyer where $\phi_{mey}(t)$ represent the inverse Fourier graph of Equation (3) and $\phi(t)$ represents the plot from Equation (9).

Figure 7. Wavelet of Meyer where $\psi_{mey}(t)$ represent the inverse Fourier graph from Equation (4) and $\psi(t)$ represents the plot of derived Equation (12).

## III. Concluding Remarks

This note offers a novel reading of the Meyer wavelet in *time domain*, which was derived by solving integrals with the aid of "tables of integrals" and rules of integration. Therefore, this approach can be used to a number of applications where closed expressions are essential. These wavelets expressions have been computed on MathCad[TM] and a graphic assessment between integral expression and the derive expressions is shown. This approach seems a natural candidate in applications involving continuous wavelet-based systems, such as rotating machine fault, wavelet-based OFDM systems, EEC signal analysis, velocity pulse signal analysis and Palmprint Based Recognition, with less computational resources to build them.